 \documentclass[%
 prl,%
 amsmath,amssymb,
preprint,%
 reprint,%
twocolumn,
]{revtex4-2}
\usepackage{xcolor}
\usepackage{graphicx}% Include figure files
\usepackage{dcolumn}% Align table columns on decimal point
\usepackage{bm}% bold math
\begin{document}

\preprint{AIP/123-QED}

\title{Record negative photoconductivity in N-polar AlGaN/GaN quantum-well heterostructures}

\author{Maciej Matys}
 \altaffiliation{Fujitsu Limited, Atsugi, Kanagawa, 243-0197, Japan}
 \email{matys.maciej@fujitsu.com}
\author{Atsushi Yamada}
\affiliation{Fujitsu Limited, Atsugi, Kanagawa, 243-0197, Japan}
\author{Toshihiro Ohki}
\affiliation{Fujitsu Limited, Atsugi, Kanagawa, 243-0197, Japan}
\author{Kouji Tsunoda}
\affiliation{Fujitsu Limited, Atsugi, Kanagawa, 243-0197, Japan}

\date{\today}% It is always \today, today,

\begin{abstract}
The AlGaN/GaN quantum-well heterostructures typically exhibit a positive photoconductivity (PPC) during the light illumination. Surprisingly, we found that introducing the GaN/AlN superlattice (SL) back barrier into N-polar AlGaN/GaN quantum-well heterostructures induces a transition in these heterostructures from PPC to negativie photoconductivity (NPC) as the SL period number increased at room temperature. This transition occurred under an infrared light illumination and can be well explained in terms of the excitation of hot electrons from the two-dimensional electron gas and subsequent trapping them in a SL structure. The NPC effect observed in N-polar AlGaN/GaN heterostructures with SL back barrier exhibits photoconductivity yield exceeding 85$\%$ and thus is the largest ones reported so far for semiconductors. In addition, NPC signal remains relatively stable at high temperatures up to 400 K. The obtained results can be interesting for the development of NPC related devices such as photoelectric logic gates, photoelectronic memory and infrared photodetectors.
\end{abstract}
\maketitle

\section{Introduction}

The photoconductivity is a phenomenon in which the electric conductivity of a material increases upon the light illumination. In the literature, this effect is often called the positive photoconductivity (PPC) and is commonly observed in various semiconductors. However, in rare cases it is possible to observe the opposite effect to PPC, i.e. negative photoconductivity (NPC), in which the electric conductivity decreases upon illumination \cite{1,2,3,4,5}. This anomalous effect was found in certain low-dimension materials, such as, nanowires,\cite{6,7,8} MoS$_2$ monolayers \cite{9}, quantum dots \cite{10}, nanorods \cite{11}, degenerated InN thin films \cite{12}, graphene \cite{13}, Van der Waals heterostructures \cite{14,15}, as well as in a few bulk materials \cite{16,17,18,19,20,21,21a} and InAs/InGaAs quantum well heterostructures \cite{22}. The NPC effect found great potential applications in optoelectronic memory, low-power photodetectors, gas sensors etc. \cite{23,24,25,26,27,28,29,30} Furthermore, the combination of NPC and PPC effects can be used to build a photoelectric logic gate \cite{31}.

\begin{figure}
\includegraphics{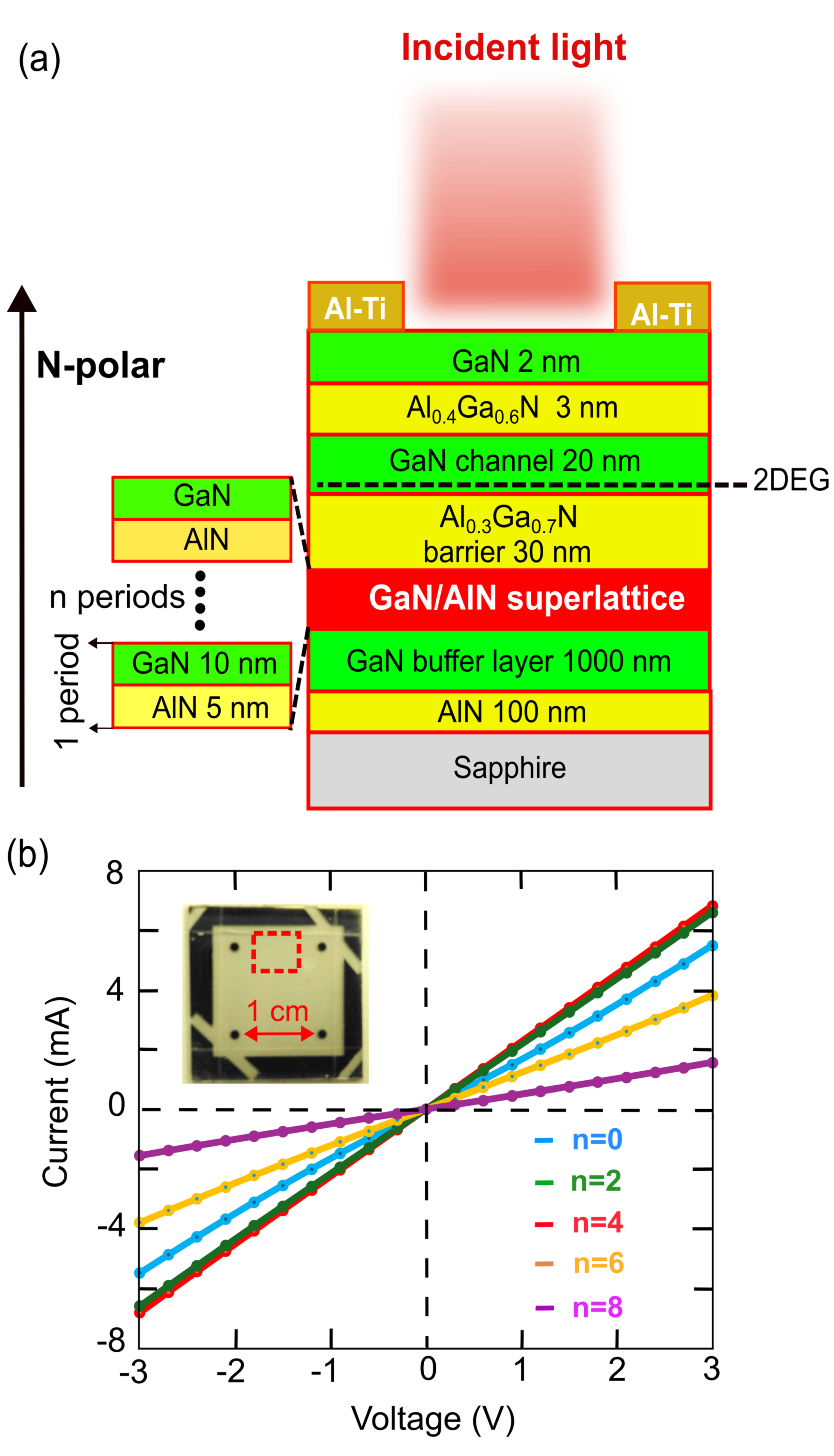}% Here is how to import EPS art
\caption{\label{fig:epsart} (a) Schematic illustration of the fabricated heterostructures used in this study. (b) Current-voltage characteristics of investigated structures. Inset of (b) shows the optical image of an actual device with marked illuminated area by the dashed red lines.}
\end{figure}

For almost 30 years, the AlGaN/GaN quantum well heterostructures were widely investigated because of their potential applications in fabricating various electrical and optical electronic devices such as field-effect transistors and photodetectors \cite{32}. However, despite this fact experimental observations of the NPC effect in these heterostructures were very rare and limited mainly to a very low temperature \cite{33}, which hinders practical development of NPC devices based on the AlGaN/GaN heterostructures. In this work, we surprisingly discovered that introducing the GaN/AlN superlattice (SL) back barrier into N-polar AlGaN/GaN heterostructures caused the transition in these heterostructures from PPC to NPC effect as the SL period number increased at the room temperature during infrared light illumination. The observed NPC effect in the N-polar AlGaN/GaN quantum-well heterostructure with SL back barrier is the largest one reported for semiconductors so far - it exhibits the photoconductivity yield exceeding 85$\%$. We proposed the explanation of this phenomenon in terms of the excitation of hot electrons from the two-dimensional electron gas (2DEG) and subsequent trapping them in the SL structure.

\section{Sample structure and experiment}

In this paper, we investigated the photoconductivity of the N-polar AlGaN/GaN heterostructures with GaN/AlN SL back barrier whose schematic illustration is shown in Fig. 1(a). The photoconductivity was analyses as a function of GaN/AlN SL periods $n$, which was varied from 0 to 8 (see inset in Fig. 1(a)). It should be emphasized that although the deposited AlN layer in SL is probably not a pure AlN but rather high Al content AlGaN layer (see later Section "Composition and structural properties") we will call here our SL as "GaN/AlN SL".

\begin{figure}
\includegraphics{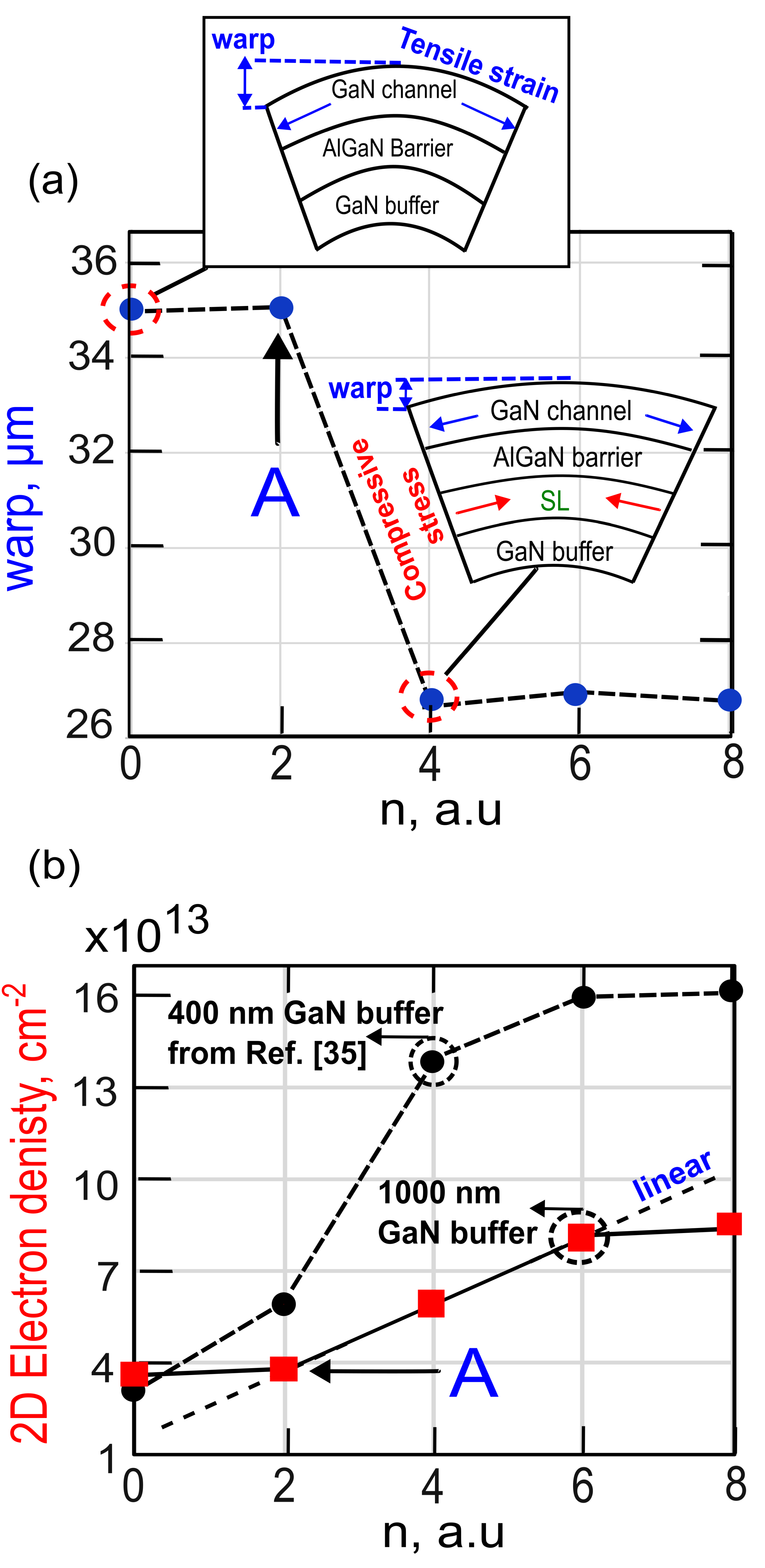}% Here is how to import EPS art
\caption{\label{fig:epsart} (a) Dependencies of warp parameter, \emph{w}, as a function of \emph{n} and (b) dependencies of 2D electron density vs. $n$. Inset in (a) shows wafer flattering phenomenon: deposition of SL caused compressive stress which reduced tensile strain. Arrow A ($n$=2) indicates the starting point of warp decreasing (a) and 2D electron density increasing (b).}
\end{figure}

The investigated structures were grown by metal-organic chemical vapor deposition (MOCVD) using trimethylgallium (TMGa), trimethylaluminum (TMAl), and ammonia as precursors on a (0001) sapphire substrate. The process started from the growth of 100 nm AlN layer at 1025 $^o$C followed by growth of 1 um unintentionally doped GaN buffer layer. Next, on the top of the GaN buffer layer a thin SL structure was deposited containing the alternating 10 nm thick GaN and 5 nm thick AlN layers as shown in the inset of Fig. 1(a). Subsequently, a 30-nm-thick Al$_{0.3}$Ga$_{0.7}$N barrier layer was deposited on the SL structure followed by the growth of 20-nm thick undoped GaN channel layer. Finally, 3-nm-thick Al$_{0.4}$Ga$_{0.6}$N and 2-nm-thick GaN cap layers were deposited on the top of the structure. 
It should be noted that the Al$_{0.4}$Ga$_{0.6}$N layer was deposited in order to minimize the capturing of electrons (from 2DEG at the GaN/Al$_{0.3}$Ga$_{0.7}$N interface) by the surface states located at the as-grown epi surface. The details of the fabrication process of superlattice structures can be found in Ref. [35]. For the photocurrent measurements, the aluminum-titanium ohmic contacts were fabricated, as shown in Fig. 1(a). The current-voltage (I-V) characteristics proved good ohmic properties of contacts in all samples with different SL period numbers (see Fig. 1 (b)). The photoconductivity experiments were performed using a 150 W halogen lamp with filters of 430 nm, 610 nm and 990 nm wavelengths at room temperature and 400 K. The light intensity was kept at a relatively low level of 40 mW/cm$^{2}$ to avoid sample heating. The illuminated area is indicated by the dashed red lines in the inset of Fig. 1 (b). 

\section{Basic properties of N-polar AlGaN/GaN heterostructures}

To understand better the mechanism of photoconductivity, we performed in this Section the detailed characterization of the electronic properties and chemical composition of our samples.

\subsection{Strains in AlGaN/GaN heterostructures with superlattice}
 
In this Section we analyzed the evolution of strains in the AlGaN/GaN heterostructures due to deposition of AlN/GaN SL. This is an important point for understanding the changes of the two dimensional (2D) electron density with $n$. To estimate roughly the strain in the fabricated samples, we measured the flattening of a wafer by determining the standardized warp and bow parameter (defined in the inset of Fig. 2(a)). In general, there is a linear relationship between the warp parameter ($w$) and residual stress ($\sigma_f$) in the film, as follows \cite{34a,34b}:
 
\begin{equation}\label{wz_2}
   \sigma_f=\frac{8wE_sh_s^2}{6d^2h_f(1-v_s)}
\end{equation}
 
where $d$ is the wafer diameter, $E$ is the Young’s modulus, $h$ is the thickness, $v$ is the Poisson’s ratio, and the subscripts $f$ and $s$ correspond to the film and substrate, respectively. Eq. 1 shows that the changes of the warp parameter, $w$, directly correspond to the changes of residual stress in the film, $\sigma_f$.
 
Fig. 2(a) shows the dependencies of the $w$ parameter as a function of the SL period number, $n$, (the period number of SL structure was defined in Fig. 1(a)). One can note from Fig. 2(a) that the $w$ parameter is a step-like function with $n$, namely below $n\leq$2, $w$ is constant and for $n$ between 2 and 4, it suddenly drops and again for $n\geq$4 becomes constant. This means that the deposition of the SL structure with $n>$2 causes flattening of the wafer, as schematically shown in the inset of Fig. 2(a). This flattening appears because of the reduced tensile strain, which occurs due to generation of the compressive stress in the SL structure (see inset of Fig. 2(a)).
 
\subsection{Electronic properties}
 
The changes of the $w$ parameter with $n$ correspond very well with changes of 2D electron density obtained from the Hall-effect measurements by the Van der Pauw method (with the magnetic field of 1 T) as shown in Fig. 2(b). From this Figure, it is clear that the 2D electron density starts to increase for the same $n$ for which the $w$ parameter suddenly decreases (Fig. 1(a)). An increase of the 2D electron gas density is due to reduction of the tensile strain, which leads to a decrease of the piezoelectric polarization charges in tensile layers (for more details see Ref. [35]). However, the observed increase of the 2D electron density with $n$ is much smaller than for the similar AlGaN/GaN heterostructure with an 400 nm thick GaN buffer layer, as shown in Fig. 2(b), which suggests that the thickness of the GaN buffer layer is important in determining the electronic properties of N-polar AlGaN/GaN heterostructures with SL structures on sapphire substrates. In addition, for $n$ from 2 to 6, the 2D electron density seems to be linearly dependent on $n$, which is also different than in the case of the 400 nm thick GaN buffer layer. The linear dependence of the 2D electron gas density vs. $n$ suggests that the parallel 2D electron gas channels were formed in the SL structure after wafer flattening, which caused an enhancement of the fixed polarization charges (due to reduction of piezoelectric polarization charges in tensile layers.

\begin{figure}
\includegraphics{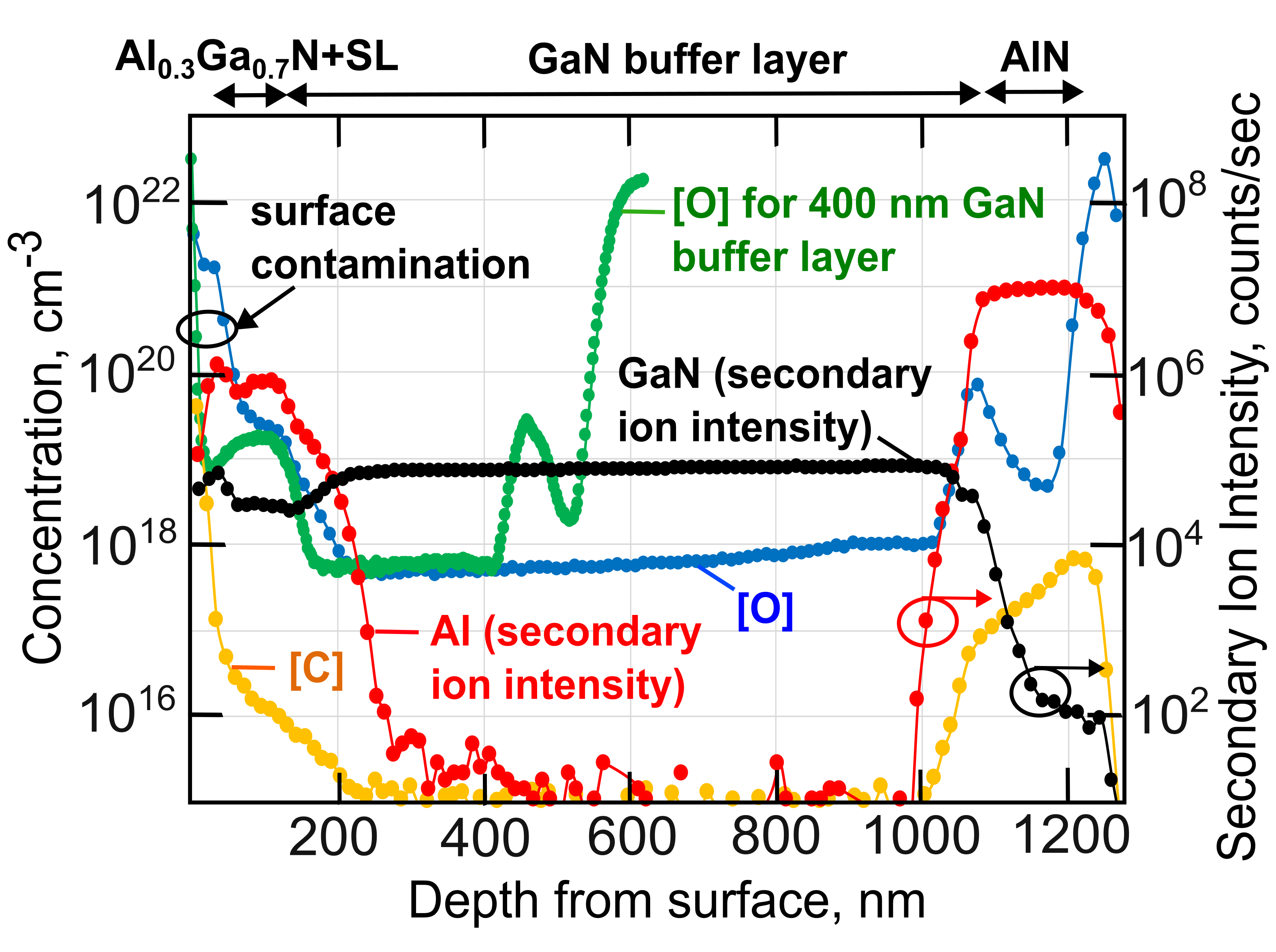}% Here is how to import EPS art
\caption{\label{fig:epsart} SIMS depth profile of investigated N-polar AlGaN/GaN heterostructure with 4 period SL and 1 $\mu$m thick GaN buffer layer. For comparison, a depth distribution of [O] for N-polar AlGaN/GaN heterostructure with 4 period SL and 400 nm thick GaN buffer layer is shown. The atomic concentration was determined for O and C.} 
\end{figure}

\subsection{Composition and structural properties}

The chemical composition analysis of the N-polar AlGaN/GaN heterostructures with SL structures was performed using the Secondary Ion Mass Spectrometry (SIMS) depth profiling, as shown in Fig. 3. At first, we noticed an elevated level of the oxygen impurity concentration ([O]): in the GaN buffer layer [O] was at the level of 4.8$\times10^{17}$cm$^{-3}$ and then increased toward the surface reaching 2$\times10^{19}$cm$^{-3}$ in the top Al$_{0.3}$Ga$_{0.7}$N layer. The incorporation of the oxygen during the growth of N polar GaN layers is always larger than in the case of Ga-polar GaN \cite{34c}. However, the [O] value of 4.8$\times10^{17}$cm$^{-3}$ in the GaN buffer layer seems to be relatively low as for N-polar GaN layers (typical values of [O] in N-polar GaN layers are $10^{18}$cm$^{-3}$ \cite{34c}). In addition, it should be noted that [O] level in the investigated AlGaN/GaN heterostructures does not change too much with the GaN buffer layer thickness, as shown in Fig. 3, which shows also the depth distribution of [O] for AlGaN/GaN heterostructures with the 400 nm thick GaN buffer layer. The high level of [O] in the top Al$_{0.3}$Ga$_{0.7}$N can explain well why for $n$=0 (without SL structure) the AlGaN/GaN heterostructures with 400 and 1000 nm thick GaN buffer layers exhibit the 2D electron density of 3-3.5$\times10^{13}$cm$^{-2}$ (see Fig. 2(b)) which is higher than that resulting only from the spontaneous and piezoelectric polarization at the GaN/Al$_{0.3}$Ga$_{0.7}$N interface. At the same time, the carbon concentration [C] was below a detection limit ($10^{16}$cm$^{-3}$) in the GaN buffer layer and above $10^{16}$cm$^{-3}$ in the SL region. Furthermore, we also found that Ga atoms are rather uniformly distributed in the SL structure (see Fig. 3) which suggests that the AlN layers in SL are not pure AlN but rather high Al content AlGaN. This explains the same 2D electron density for $n$=0 and $n$=2 (see Fig. 2(b)). Namely, as we found previously \cite{34} from the band diagram simulation, the additional 2D electron gases in the SL structure should not appear if the Al$_{0.6}$Ga$_{0.4}$N layer is present in SL instead of the AlN layer. In other words, the 2D electron density in the AlGaN/GaN heterostructure with Al$_{0.6}$Ga$_{0.4}$NGaN SL should be the same as in the AlGaN/GaN heterostructure without SL. 
 
\begin{figure}
\includegraphics{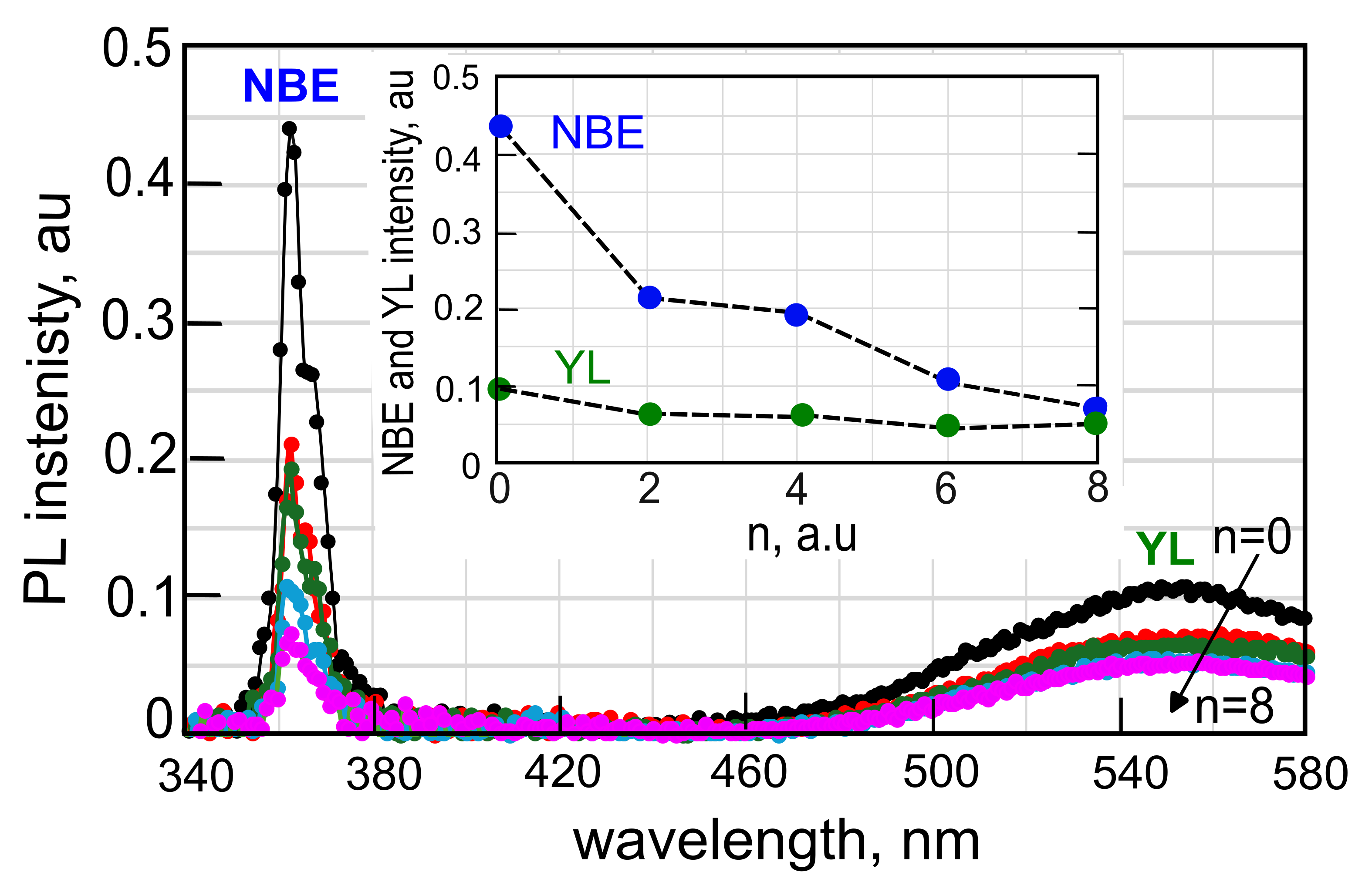}% Here is how to import EPS art
\caption{\label{fig:epsart} Room temperature PL spectra of investigated N-polar AlGaN/GaN heterostructures with SL structures ($n$ varied from 0 to 8). Inset compares the intensity of near-band emission (NBE) and yellow emission (YL) peaks vs. $n$.}
\end{figure}

\subsection{Crystalline quality}
 
To check the crystalline quality of the investigated AlGaN/GaN heterostructures, we carried out the photoluminescence (PL) measurements at room temperature. Fig. 4 shows the PL spectra in the 340-580 nm range from the AlGaN/GaN heterostructures with $n$ from 0 to 8. In these PL spectra, we can distinguish two main PL peaks, namely one peak located at 362.1 nm is related mainly to the band-to band recombination (and perhaps to free excitons) in GaN layers and is called the near band edge emission (NBE) peak. The second one located at 550.3 nm is related to so-called yellow luminescence (YL). As can be seen, the highest NBE peak was observed for $n$=0 and lowest for $n$=6 and 8. With increasing $n$, the NBE peak gradually decreases, as shown in the inset of Fig. 4. Simultaneously, the YL peak was practically constant with $n$. This clearly shows that the non-radiative defect states are responsible for decreasing of the NBE peak with $n$. Increasing of the number of non-radiative defect states with $n$ is a direct indicator that crystalline quality of SL becomes worse with $n$.

\section{Results and Discussion} 

\subsection{Photocurrent data} 

\begin{figure}
\includegraphics{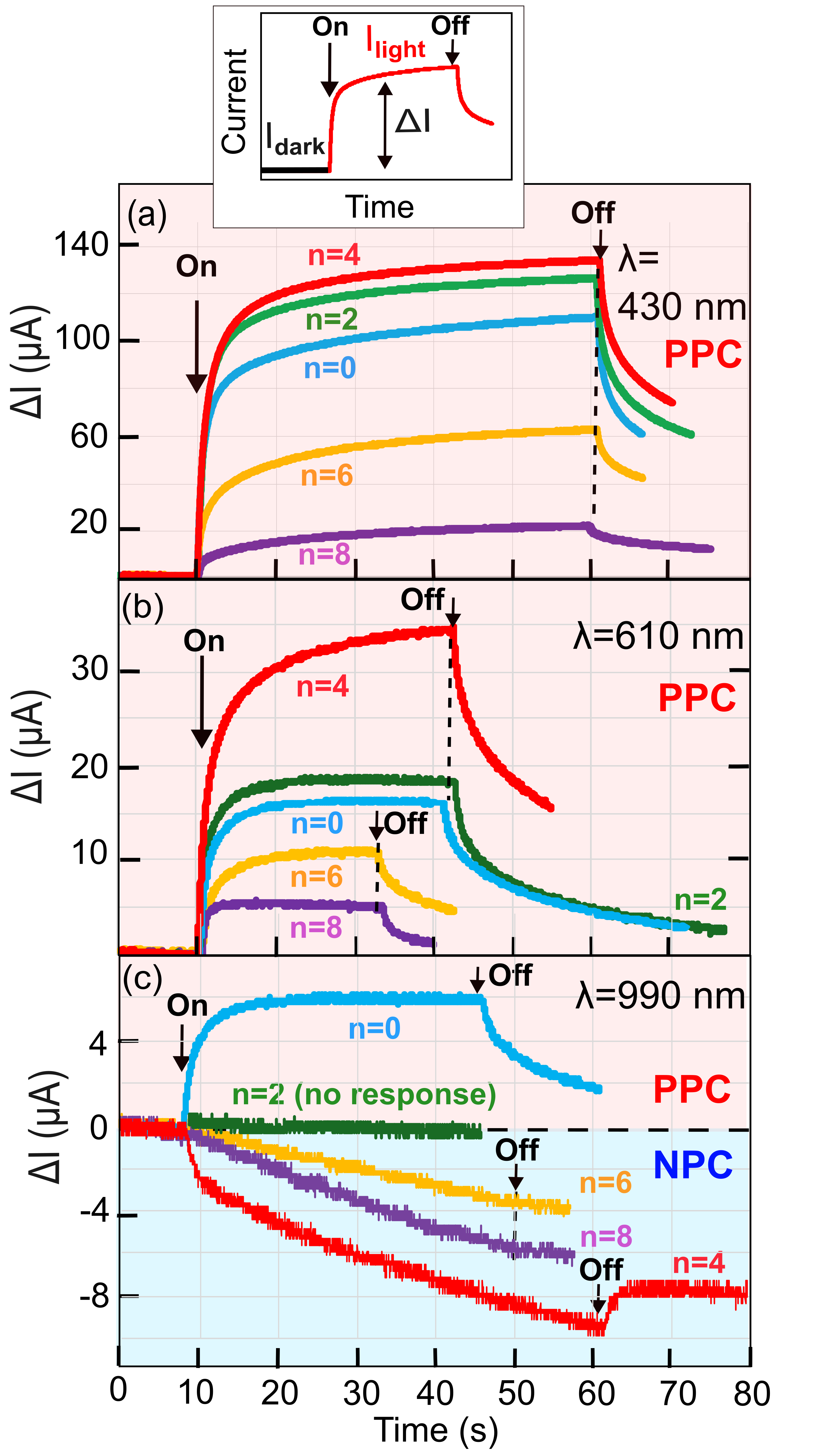}% Here is how to import EPS art
\caption{\label{fig:epsart} The transient photocurrent characteristics of N-polar AlGaN/GaN heterostructures with SL period $n$=0, 2, 4, 6 and 8 obtained under the illumination with wavelengths: (a) 430 nm, (b) 610 and (c) 990 nm. Inset of (a) shows schematically definition of the photocurrent $\Delta I$.}
\end{figure}

Fig. 5 presents the transient photocurrent ($\Delta I$) measurements (light on/off) performed for N-polar AlGaN/GaN heterostructures with SL period $n$=0, 2, 4, 6 and 8 at room-temperature under a bias voltage of 2 V and illumination with wavelengths: (a) 430 nm, (b) 610 nm and (c) 990 nm. The photocurrent, $\Delta I$, is defined as the difference between the current under illumination ($I_{light}$) and in the dark ($I_{dark}$) as schematically shown in the inset of Fig. 5(a). 

In the case of the visible light illumination (Figs. 5(a) and (b)) all samples exhibited the PPC effect. However, in the case of the infrared illumination (Fig. 5(c)) a clear transition from PPC to NPC was observed with SL period increasing. In particular, the PPC effect occurred for the sample with $n$=0 whereas for the samples with $n\geq4$, the current decreased after the infrared light turn-on that indicated the NPC effect. It is interesting to note that for the sample with $n$=2 there was no response at the infrared light illumination. In addition, for the sample with $n$=4 the current suddenly decreased (after turn on the light) but for the samples with $n$=6 and 8 this decreasing was less rapid. The strongest NPC effect, i.e. the largest current drop due to the illumination was observed for $n$=4. For all samples where the NPC effect occurred, $\Delta I$ was far from the saturation within 60 s. On the other hand, in PPC, $\Delta I$ was saturated within 60 s for almost all cases. Furthermore, for the samples exhibiting NPC, the long persistent photoconductivity was observed, i.e. the current needed a long time to return to its initial dark value after the light turn-off. However, for $n$=4 after switching off the light, the photoconductivity drops more rapidly than in the cases of $n$=6 and $n$=8 where it is difficult to distinguish the moment of switching off illumination (on and off photocurrent signal is practically the same). One of the possible explanations of this discrepancy can be a sudden escape of some trapped electrons from SL to the GaN buffer layer by, for example, the tunneling effect which can be more significant for SL with $n$=4 than for $n$=6 and 8 because of the lower barrier thickness (tunneling effect becomes weaker with the increasing barrier thickness). The long persistent NPC effect can be interesting for applications in the NPC optical memory devices. Contrary to NPC, the PPC decay is much faster after switching off the light (see Figs. 5(a)-(b) and (c) for samples with $n$=0). This observation is consistent with the previous reports in which PPC decayed much faster than NPC \cite{21}.

\begin{figure}
\includegraphics{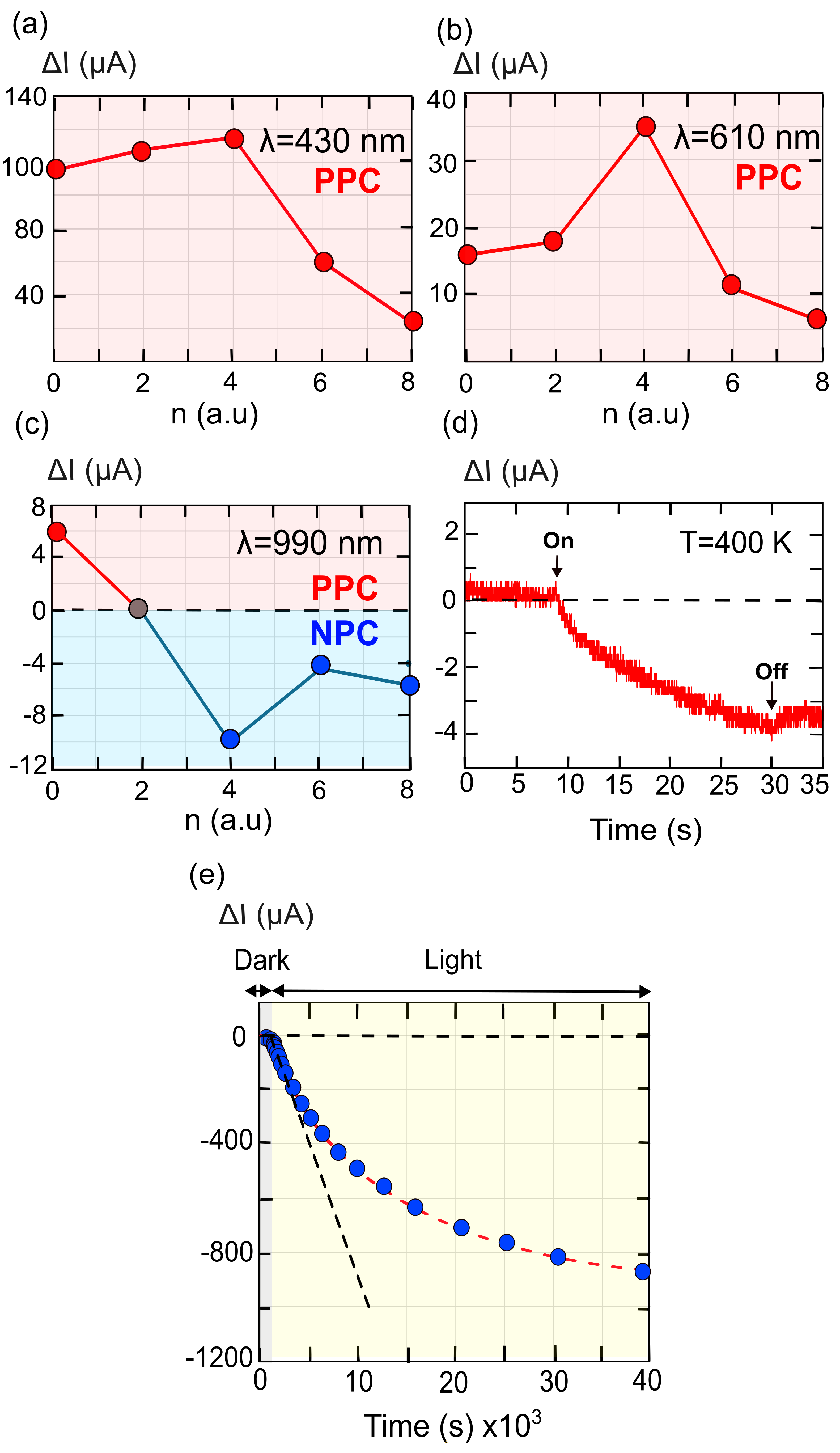}% Here is how to import EPS art
\caption{\label{fig:epsart} Dependencies of $\Delta I$ from Fig. 2 as a function of the SL period number $n$ in the case of: (a) 430 nm, (b) 610 and (c) 990 nm. (d) Transient photocurrent characteristics of N-polar AlGaN/GaN heterostructures with n=4 at 400 K under the infrared light illumination. (e) Transient photocurrent characteristics of N-polar AlGaN/GaN heterostructures with n=8 at room-temperature under the infrared light illumination for 40000 s.}
\end{figure}

In Figs. 6(a)-(c), we summarized the dependencies of the photocurrent from Fig. 5 as a function of the SL period number $n$. In the case of 430 nm (Fig. 6(a)) $\Delta I$ weakly depends on $n$ up to $n$=4 above which $\Delta I$ decreases. The similar trend was also observed for a wavelength of 610 nm (Fig. 6(b)), except for $n$=4 for which an enhancement of $\Delta I$ was registered. The evident decrease of $\Delta I$ for $n>$ 4 excellently correlates with a decrease of the NBE peak vs. $n$ (see Fig. 4). From these observations, we can conclude that non-radiative defect states are the reason of the $\Delta$I decrease above $n>$4 for 430 nm and 610 nm (Figs. 6(a) and (b)). However it is still unclear why for 610 nm at $n$=4 $\Delta$I was enhancement (Fig. 6(d)). Under the infrared light illumination (Fig. 6(c)) $\Delta I$ changed significantly for the samples between $n$=0 and $n$=4 with a clear change of sign at $n$=2 from the positive to negative value. Above $n$=4, it seems that $\Delta I$ become saturated with $n$. For the potential applications in the infrared photodetectors, we also verified the temperature impact on the observed NPC effect, as shown in Fig. 6(d). As can be seen from this figure, the temperature rising up to 400 K did not cause a significant decrease of NPC effect compared to the room temperature (see Fig. 5(c) curve for $n$=4). We can even say that the obtained NPC effect was relatively stable with temperature which is advantageous for applications in NPC devices. For the estimation of magnitude of the measured NPC, we calculated the photoconductivity yield defined as \cite{21}:

\begin{equation}
Y_{PC}=\frac{I_s-I_{dark}}{I_{dark}}
\end{equation}

where $I_s$ is the saturation current (after illumination). Because in our case NPC was not saturated within 60 s (see Fig. 5(c)), we extended the photocurrent transient measurements to much longer times, as show in Fig. 6(e). For this experiment (i.e. estimation of maximal photoconductivity yield) we have selected the sample with $n$=8 since this sample exhibits the lowest $I_{dark}$ (see Fig. 1(b)). It is evident from Fig. 6(e) that $\Delta I$ starts to saturate around 5000 s but the full saturation is not reached within 40000 s. Nevertheless, this figure clearly indicates that the registered NPC effect is huge. In particular, at 40000 s, $\Delta I\approx$ 0.85 mA, which is 85$\%$ of the dark current $I_{dark}$= 1 mA (for sample with $n$=8, see Fig. 1(b) at bias 2 V). Therefore, we can assume that $Y_{PC}$ is $>$ 85$\%$ (see Eq. 2). For comparison, in the case of graphene $Y_{PC}$ of only 34$\%$ was reported. Note that the authors registered the full saturation of the photo-current after 7500 s. After this time the authors observed the current decreasing by 34$\%$ \cite{13}. The highest $Y_{PC}$ was observed so far for Bi-doped MAPbBr$_{3}$ perovskites. In this case, the photo-current was almost saturated after 1000 s illumination and $Y_{PC}$ of 67$\%$ was obtained \cite{21}. Our $Y_{PC}$ significantly exceeds this value, which means that it is the highest ever reported for semiconductors. In this context, it should be highlighted that the large NPC effect with the value of 50-90$\%$ was also observed in the Van der Waals heterostructures \cite{14,15}. However, in those structures the high $Y_{PC}$ value $>$90$\%$ was achieved due to using the floating gate structure but not as an intrinsic material property \cite{14}. Thus, we believe that our NPC effect is the highest one obtained for semiconductor materials.

\subsection{Discussion} 
\begin{figure}
\includegraphics{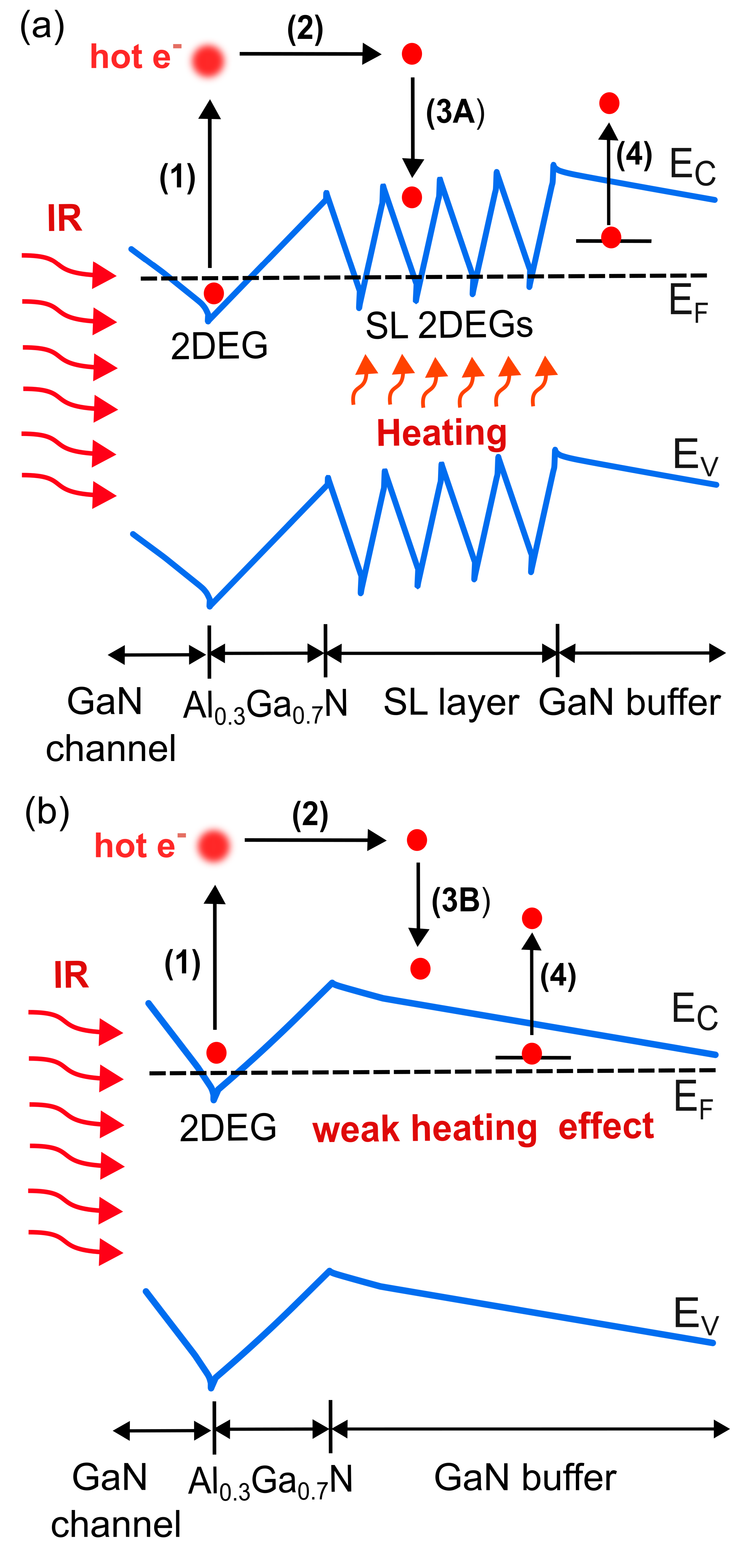}% Here is how to import EPS art
\caption{\label{fig:epsart} Possible band diagrams of investigated N-polar AlGaN/GaN heterostructures: (a) with SL back barrier and (b) without SL back barrier. The marked processes on (a) and (b) mean: (1) excitation of hot electrons from quantum well, (2) diffusion of hot electrons, (3A) thermalization of hot electrons in the SL structure, (3B) thermalization of hot electrons in the GaN buffer layer and (4) excitation of electrons from deep-levels in a GaN buffer layer.}
\end{figure}

The origin of NPC phenomenon was attributed previously to generation of hot electrons, formation of metastable states (in particular the light induced trap states and DX centers), intraband transitions, surface plasmon scattering, trapping by deep-levels, adsorption of moisture and trion formation \cite{9,10,21,22,35,36}. From all of these hypotheses, the hot electron-based ones seem to be most suitable for understanding the occurrence of NPC effect in our AlGaN/GaN samples (just like it was observed in 2DEG InAs/InGaAs systems \cite{22}). Below we will show that the observed NPC effect can be well explained in terms of hot electron generation and thermalization. 

From the SIMS depth profile analysis (see Fig. 3) we found that the fabricated GaN/AlN SL may not contain the pure AlN layer but rather high Al-content AlGaN layer. On the other hand, such layer exhibits much lower thermal conductivity than pure AlN or GaN because of the dominant phonon-alloy scattering \cite{37}. Thus, introducing the SL back barrier structure to the investigated N-polar AlGaN/GaN heterostructure should lead to reducing the thermal conductivity. In other words, this means that the samples with the SL structure can be heated more effectively than the samples without SL. Next, upon the infrared radiation the hot electrons are excited from the quantum well at GaN/Al$_{0.3}$Ga$_{0.7}$N interface, as schematically shown in Fig. 7(a) (process 1). Although the dependence of 2D electron density vs. $n$ (see Fig. 2(b)) suggested the formation of parallel 2DEG channels in SL, in the case of $n>$ 2 for simplicity we assumed that the infrared light causes excitation of hot electrons only from one 2DEG located at the GaN/Al$_{0.3}$Ga$_{0.7}$N interface. This assumption, as we show later, was well supported by experimental observations. The excited hot electrons from the quantum well are subsequently trapped in the SL structure (see Fig. 7(a) process 3A) which causes that the energy of hot electrons is transferred to the host lattice leading to heating the near GaN/Al$_{0.3}$Ga$_{0.7}$ interface region due to the low thermal conductivity of the SL structure and thereby reducing the electron mobility in the 2DEG channels. Beside the excitation electrons from the 2DEG at GaN/Al$_{0.3}$Ga$_{0.7}$N interface, the infrared light must also lead to the excitation of electrons from deep-levels in the GaN buffer layer, as shown in Fig. 7 (process 4), since a relatively high PPC effect was observed in the structure without SL (see Fig. 5). 

Based on the above processes, we can explain the observed NPC effect using the following approach. Firstly, we defined the dark current density, $I_{dark}$ in our heterostructures. The main conductivity channels are 2DEGs located at the GaN/Al$_{0.3}$Ga$_{0.7}$N interface and 2DEGs in the SL structure as it was suggested by linear dependencies of the 2D electron density vs. $n$ (see Fig. 2(b)). However, some contribution to the conductivity may also comes from the bulk electrons in the GaN buffer layer. Thus, $I_{dark}$ is expressed as follows:

\begin{equation}
I_{dark}=q \mu_{2DEG}^{D}\sum_{i=1}^{i=n} n_i+qn_b\mu_{b}
\end{equation}

where $n_i$ is the 2DEG concentration in the ith channel ($i$=1 corresponds to the channel at the GaN/Al$_{0.3}$Ga$_{0.7}$N/ interface), $\mu_{2DEG}^{D}$ is the 2DEG mobility in the dark (we roughly assumed that the electron mobility is the same in the each 2DEG channel), $n_b$ is the concentration of electrons in the bulk and $\mu_{b}$ is the bulk electron mobility.

As we explained before, under the infrared illumination two processes occur: (\emph{i}) the electrons from the deep levels in the GaN buffer layer are excited to the conduction band and (\emph{ii}) 2DEG mobility is reduced due to the hot electron excitation and their subsequent trapping in SL (see process 1 and 3A in Fig. 7(a)). These processes modified the current as follows:

\begin{equation}
I_{light}=qn_s\mu_{2DEG}^{L}\sum_{i=1}^{i=n} n_i+q(n_b+\Delta n)\mu_{b}
\end{equation}

where $\mu_{2DEG}^{L}$ is the 2DEG mobility in the case of light illumination and $\Delta n$ is the concentration of excess electrons excited from the deep levels in GaN buffer layer. Note that in Eq. 4 we reasonably assumed that the number of excited hot electrons is negligible compared to $\Delta n$. From the combination of Eq. 3 and 4, we obtain the expression for the photocurrent $\Delta I$ as:

\begin{equation}
\Delta I=q (\mu_{2DEG}^{L}-\mu_{2DEG}^{D}) \sum_{i=1}^{i=n} n_i+q\Delta n\mu_{b}
\end{equation}

\begin{figure}
\includegraphics{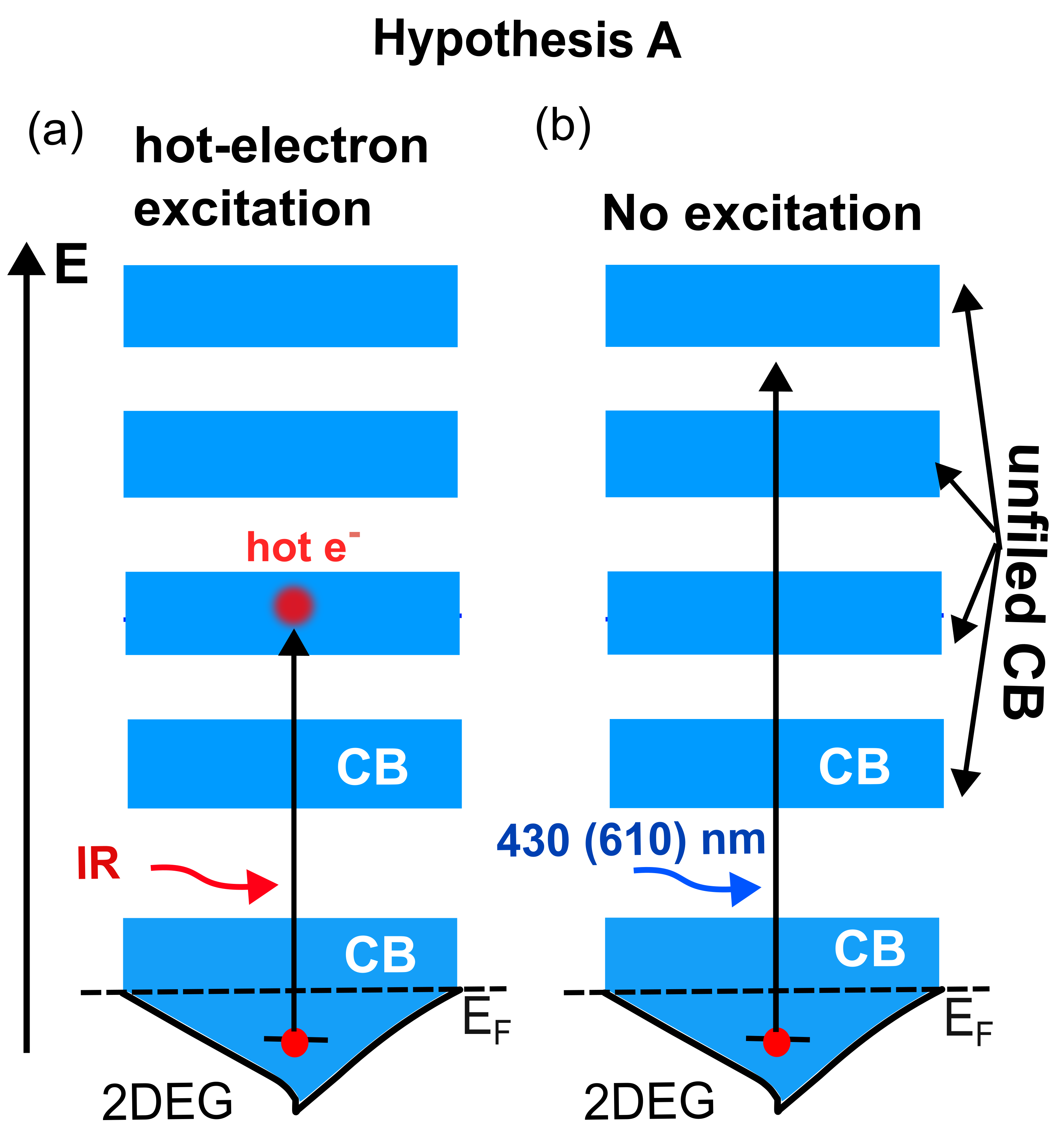}% Here is how to import EPS art
\caption{\label{fig:epsart} Schematic illustration of photoexcitation of hot electrons according to the Stark-Einstein rule (hypothesis A).}
\end{figure}

The trapped hot electrons in the SL structure (process 3A in Fig. 7(a)) lead to heating the near GaN/Al$_{0.3}$Ga$_{0.7}$N/ interface region due to low thermal conductivity of SL and reducing 2DEG mobility in the channels. Then, if this reduction is sufficiently high, the following term $\mid{q(\mu_{2DEG}^{L}-\mu_{2DEG}^{D}) \sum_{i=1}^{i=n} n_i}\mid$ in Eq. 5 will be larger than $q\Delta n\mu_{b}$ leading to the negative photocurrent ($\Delta I <$ 0) and thus the NPC effect. In the case of the sample without SL, the excited hot electrons from the quantum well are relaxed in the GaN buffer layer, as schematically shown in Fig. 7(b) (process 3B). This process does not lead to heating the near GaN/Al$_{0.3}$Ga$_{0.7}$N interface region since the thick GaN buffer layer exhibits high thermal conductivity. Therefore, in the sample without SL $\mu_{2DEG}^{L}\approx\mu_{2DEG}^{D}$ and thus according to Eq. 5, we observed in this sample the PPC effect equal to $q\Delta n\mu_{b}$. 

The hot electron model can also explain in detail the observed dependencies of $\Delta$I as a function of $n$ in the case of 990 nm (Figs. 6(c)) (especially transition from PPC to NPC). Namely, after the excitation from 2DEG (process 1, Fig. 7(a)) due to the infrared light illumination the hot electrons diffuse toward GaN buffer layer (process 2, Fig. 7(a)). Next, if the length of the SL structure is shorter than the diffusion length of hot electrons, the excited carriers will not be trapped in SL but they reach the GaN buffer layer where they thermalize. As it was explained before, the thermalization of hot electrons in the GaN buffer layer will not lead to heating near the GaN/Al$_{0.3}$Ga$_{0.7}$N interface region and reducing electron mobility in the 2DEG channels because the thick GaN buffer layer exhibits high thermal conductivity. Therefore, for the sample with a short SL, we should observe the PPC effect similar like in the case of samples without SL. In our case, this situation could occur for SL with $n$=1 (see Fig. 6(c)). On the other hand, while the SL length increases (i.e. the number of SL periods, $n$, rises) up to around the diffusion length of hot electrons, the part of the excited carriers will be thermalized in SL leading to heating near the GaN/Al$_{0.3}$Ga$_{0.7}$N interface region and thus decreasing of mobility in the 2DEG channels (other hot electrons will be thermalized in the GaN buffer layer. However, this effect would be too weak to overcome the PPC effect resulting from the excitation of electrons from the GaN buffer layer. In consequence, we should observe a weak PPC effect or no photocurrent signal like in the case of $n$=2 (Fig. 6(c)). Further increasing of the SL length above the diffusion length of hot electrons would lead to the situation when all excited carriers from 2DEG will be trapped in SL. This will cause a significant heating near the GaN/Al$_{0.3}$Ga$_{0.7}$N interface region and markedly reduces electron mobility in 2DEG channels causing the NPC effect as that observed for $n$=4 (Fig. 6(c)). It is also interesting to note that if we assume that hot electrons are excited only from the first 2DEG (located at the GaN/Al$_{0.3}$Ga$_{0.7}$N interface), the NPC magnitude should not grow any more vs. SL length but rather saturate with $n$ like for $n>$4 (see Fig. 6(c)) since the number of excited hot electrons is finite. In the case of excitation of hot electrons from the multiple 2DEGs, the NPC effect should increase for $n>$4 since more hot electrons are added to the system from the parallel 2DEGs channels.

We should also briefly mention that the hot-electron model proposed here can be verified by measuring the Hall effect during the light illumination. Such experiment could show if the electron mobility decreases under infrared illumination. Nevertheless, the extraction of the mobility value under the light illumination from the Hall effect is difficult task and until now only a few approaches were developed like the carrier-resolved photo-Hall technique \cite{37a}. According to Gunawan, O et al. \cite{37a} this technique allows to determine the carrier mobilities as a function of the light intensity. The application of this measurement is beyond the scope of this paper, but it is worth to consider similar experiments in the future.

The last issue is to clarify why the NPC effect was not observed under visible light illumination (Figs. 5(a) and (b)). This can be explained by the two hypothesis (here called A and B). The first hypothesis A based on the Stark-Einstein rule according to which transitions of electrons from the ground states to excited ones occur only if the energy of the incident photons match exactly the energy difference between these states. In our case, hot electrons appear due to electron excitations from the energy levels in 2DEG to unfilled bands in the conduction band as shown schematically in Fig. 8(a). The visible photons with specific wavelengths of 430 nm  and 610 nm do not much well to such energy differences and may not lead to hot electron generation (see Fig. 8(b)). Instead of that all photons with the wavelengths of 430 nm and 610 nm are mostly absorbed at the deep-levels in the GaN buffer layer. This explains why under visible light illumination a relatively strong PPC signal was observed (Figs. 5(a) and (b)).

According to the hypothesis B, both visible and infrared light leads to the generation of hot-electrons from 2DEG. However, the optical cross section for excitations of electrons from the deep-level centers corresponding to the visible light spectrum is much higher than for excitations of electrons from 2DEG region. In consequence, most of the photons with wavelengths of 430 nm and 610 nm are absorbed by the deep-levels in the GaN layer and only few lead to the generation of hot-electrons from 2DEG, which causes that the PPC effect dominates (Figs. 5(a) and (b)). In contrast, for the deep-level centers corresponding to the infrared light spectrum the optical cross section can be relatively low which causes that most of infrared photons are absorbed in the 2DEG region leading to hot-electron excitation and the NPC effect (Fig. 5(c)).

\subsection{Conclusions} 

We have experimentally observed  the negative photoconductivity effect in the N-polar AlGaN/GaN quantum-well heterostructures after introduction a GaN/AlN superlattice back barrier. The NPC phenomenon appeared in the N-polar AlGaN/GaN quantum-well heterostructures with the SL back barrier under infrared light illumination at room-temperature. The obtained NPC exhibits the photoconductivity yield exceeding 85$\%$ which is the largest value reported so far for semiconductor materials. In addition, it remains relatively stable at high temperatures up to 400 K. We show that the observed NPC can be well explained in terms of the excitation of hot electrons from the 2DEG quantum well and subsequent trapping them in the SL structure. Our finding may offer a novel approach to design NPC devices based on the AlGaN/GaN quantum well heterostructures.

\begin{acknowledgments}
The authors expresses gratitude to Norikazu Nakamura for his kind support and discussions The authors would like to thank Yoshiharu Kinoue and Takaaki Sakuyama for their support in experiments. 
\end{acknowledgments}

%\bibliography{aipsamp}% Produces the bibliography via BibTeX.

\end{document}